\documentclass[a4paper]{article}

\usepackage[english]{babel}
\usepackage[utf8x]{inputenc}
\usepackage{amsmath}
\usepackage{graphicx}
\usepackage[colorinlistoftodos]{todonotes}
 
\usepackage{caption}
\usepackage{subcaption}

\def \COUNT {$Count$}
\def \ALE {$ALEml\_undated$}

\title{Genome-scale phylogenetic analysis finds extensive gene transfer among Fungi}
\author{Gergely J. Sz\"oll\H{o}si$^1$, Adri\'{a}n Arellano Dav\'{i}n$^2$, Eric Tannier$^{2,3}$, \\Vincent Daubin$^{2,4}$, Bastien Boussau$^{2,4}$}

\newpage

\begin{document}
%\linenumbers

\maketitle

\noindent {
$^1$ELTE-MTA ``Lend\"ulet'' Biophysics Research Group, P\'azm\'any
P. stny. 1A., 1117 Budapest, Hungary;\\
$^2$Laboratoire de Biom\'etrie et Biologie Evolutive, Universit\'e de Lyon, F-69000 Lyon, France;\\
$^3$Institut National de Recherche en Informatique et en Automatique Rh\^one-Alpes, F-38334 Montbonnot, France;\\
$^4$Centre National de la Recherche Scientifique, Unit\'e Mixte de Recherche 5558, Universit\'e Lyon 1, F-69622 Villeurbanne, France;}\\

\medskip
\noindent{\bf Corresponding author:} Bastien Boussau, Laboratoire de Biom\'etrie et Biologie Evolutive, Centre National de la Recherche Scientifique, Unit\'e Mixte de Recherche 5558, Universit\'e Lyon 1, F-69622 Villeurbanne, France; Universit\'e de Lyon, F-69000 Lyon, France;  E-mail: bastien.boussau@univ-lyon1.fr\\

\begin{abstract}

Although the role of lateral gene transfer is well recognized in the evolution of bacteria, it is generally assumed that it has had less influence among eukaryotes. To explore this hypothesis we compare the dynamics of genome evolution in two groups of organisms: Cyanobacteria and Fungi.  Ancestral genomes are inferred in both clades using two types of methods.  First, \COUNT, a gene tree unaware method that models gene duplications, gains and losses to explain the observed numbers of genes present in a genome. Second, ALE, a more recent gene tree-aware method that reconciles gene trees with a species tree using a model of gene duplication, loss, and transfer. We compare their merits and their ability to quantify the role of transfers, and assess the impact of taxonomic sampling on their inferences.  We present what we believe is compelling  evidence that gene transfer plays a significant role in the evolution of Fungi.

\end{abstract}

\section{Introduction}

Reconstructing genome evolution and ancestral genomes is instrumental to  understanding the diversification of life on Earth. Doing so requires harnessing the information available in complete genome sequences, which is best achieved in a statistical framework. Integrative methods to reconstruct the evolution of genomes and thus ancestral genomes are now able to model particular histories of genes inside a general history of genomes and can integrate many different types of events. They integrate sequence-level events such as substitutions, gene-level events such as duplications (D), losses (L), and exchanges of genes between genomes, modeled by lateral gene transfers (hereafter transfers, T), as well as genome-level events such as speciations (S). This inclusiveness enables them to handle diverse groups of organisms, each with their idiosyncratic way of evolving. It therefore becomes possible to apply a single method to groups from different domains of life, and compare their modes of evolution.

Reconstructing ancestral genomes requires \textit{a minima} two types of data: extant genomes with homology relationships between genome fragments, and a tree along which these genomes are supposed to have evolved. A species tree modeling vertical descent is indispensable, because without it, we cannot differentiate vertical inheritance from lateral transfer, and little can be learned about the processes of genome evolution.

%This effect is visible for instance in a series of recent articles in which ancestral genomes were interpreted as having acquired massive amount of genes by gene transfer due to a lack of consideration of the history of species 

Using a common species tree does not mean that we assume that all homologous fragments have had the exact same history. Instead, the history of each individual homologous fragment is reconstructed, with its own succession of duplications, losses and transfers. For species that have diverged a long time ago, only the protein coding portion of the genomes is analyzed, and individual histories are reconstructed for each gene family. These gene histories are subsequently analyzed together to gain insight into genome evolution, and infer large-scale patterns of gene duplications, losses, or transfers.
Both steps, first gene tree reconstruction and second aggregation of gene histories into coherent patterns, necessitate thoughtful methodologies to overcome possible sources of errors and uncertainties.

\subsection{Reconstruction of gene histories}
\begin{figure}
\centering
\includegraphics[width=1\textwidth]{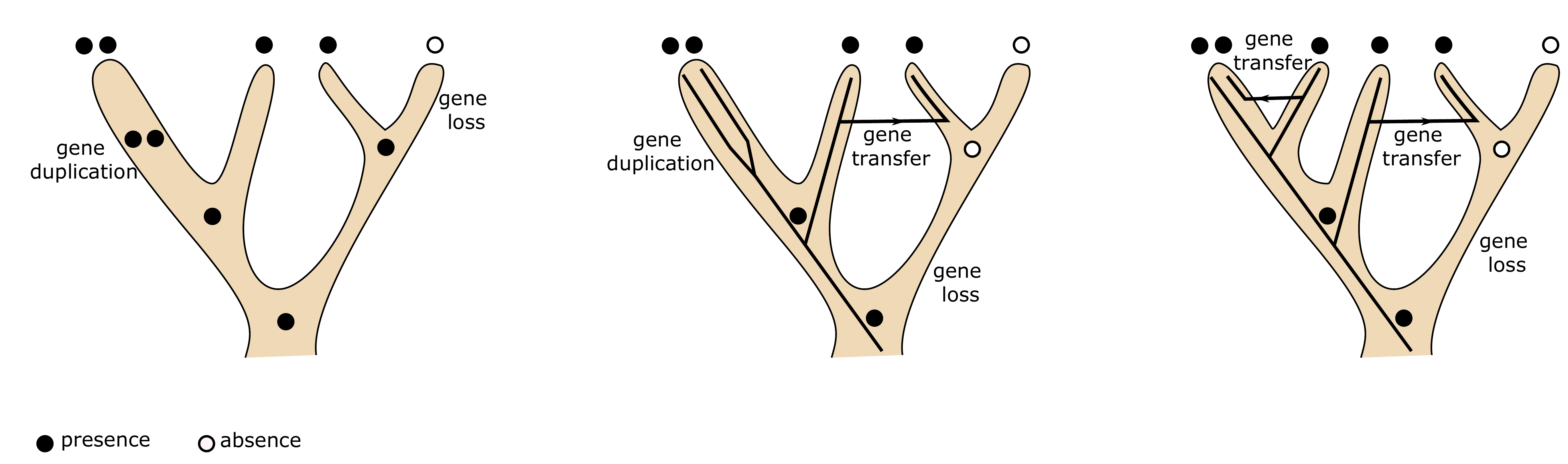}
\caption{\label{fig:2Approaches}How a gene history can be incorrectly reconstructed if the gene tree is not taken into account, or if taxonomic sampling is incomplete. Left: Inference according to the gene content approach. Middle: Inference according to a gene tree-aware approach. Right: Inference according to a gene tree-aware approach, with a more complete taxonomic sampling. Ignoring gene phylogeny and having insufficient species sampling lead to underestimation of gene transfer. In all cases, the same true gene history is assumed, but only with sufficient taxonomic sampling and with a gene tree-aware approach can it be recovered.}
\end{figure}

Gene sequences are often too short to contain sufficient information for accurate and robust reconstruction of the history of a gene family; worse, even when this information is present, models of sequence evolution may fail to capture it correctly. In general, a gene family's history cannot be reliably inferred, nor interpreted in terms of gene-level events from the set of sequences alone \cite{Szollosi2014,Groussin2014}. Using additional information coming from the species tree is a way to improve gene tree quality (Fig.\ \ref{fig:2Approaches}). This is the approach taken by "Gene tree-aware approaches". Alternatively, it is possible to entirely do away with the sequences and avoid gene tree reconstruction: the gene tree unaware ''Gene content approach" considers only gene presence/absence patterns, or numbers of genes per species. 
%Oppositely integrative  take advantage both of sequence and content, looks beyond individual gene families for additional sources of information to better reconstruct gene histories, namely other gene families.

\subsubsection{Gene content approaches}
Gene content approaches work with data in the form of either presence/absence of a gene family inside a given genome, or numbers of genes of a gene family inside a given genome (Fig.\ \ref{fig:2Approaches}). In both cases, parsimony approaches or probabilistic models have been used to reconstruct the evolution of gene families along a species phylogeny.

Among parsimony methods, one can choose between Wagner and Dollo parsimony. Choosing Dollo parsimony amounts to making a strong assumption about the pattern of gene family evolution, as it means that a gene family can be gained only once, on a single branch of the species phylogeny. In short, this means that gene transfers are forbidden. Wagner parsimony can be more moderate in its assumptions, but still requires that costs be defined for all types of events involved in gene family evolution, \textit{i.e.} duplications, transfers, losses. There is no objective way to set these costs, and users often try a range of costs, eyeball the results, and choose the costs that produce the evolutionary scenarios that seem most reasonable \cite{Boussau2004}. The most systematic approaches use ancestral genome sizes to pick costs that generate ancestral genomes that are neither too big nor too small, but still lack a proper statistical framework \cite{Dagan2007,David2010}.

Probabilistic approaches either rely on an \emph{ad hoc} adaptation of substitution models used to describe sequence evolution \cite{O2010}, or rely on a birth-death model that includes rates of gene duplication, transfer, and loss \cite{Csuros2006,Szollosi2012a}. They can include corrections for unobserved data, \textit{i.e.} gene families that are present in none of the sampled species but that were present in ancestral species \cite{O2010}. These approaches do not require arbitrary choices of costs: instead rates are estimated from the data. Different models can then be tested against each other, for instance to test whether there is significant support for the presence of gene transfer in the data. These tests rely on the well-known machinery for model testing, and include likelihood ratio tests, Akaike or Bayesian Information Criterion, or Bayes factors if inference is performed in a Bayesian setting.

% I think it is risky to talk about nelson sathi in this way without further demonstration. I added a few words at the beginning of the section to evacuate this problem.
%Recently, a new approach has been proposed to study one aspect of genome evolution \cite{Nelson-Sathi2012,Nelson-Sathi2014}. Based on gene presence-absence patterns, it aims at reconstructing all gene transfers arriving at the root of a pre-defined clade. Oddly, it assumes that all genes present in as few as 2 genomes in the clade of interest but absent in sister clades must have been acquired at the root of the said clade by transfer from a distant genome. This approach is severely biased, as by construction it overestimates gene content at the root of the clade of interest by moving all gene originations from within the clade to its root, and because it makes no attempt at orienting gene transfers, which are equally likely to arrive to or depart from the root of the clade. Because of its many flaws, this method will no longer be discussed in this article.

Whether they are analyzed by parsimony or probabilistic approaches, gene content data are limited in their ability to detect events of gene family evolution. Even the approaches that use the numbers of genes and not just their pattern of presence/absence will make mistakes that approaches based on the consideration of gene tree topologies could avoid, if the gene trees are accurately reconstructed (see middle graph in Fig.\ \ref{fig:2Approaches}).

\subsubsection{Gene tree-aware approaches}
Most gene families share parts of their histories, \textit{i.e.} have been inherited together from ancestors to descendants during parts of their history (Fig.\ \ref{fig:2Approaches}). If we can reconstruct the parts of their histories where genes have co-evolved, then jointly reconstructing gene histories can be very helpful, because more information is available to reconstruct each gene history. In cases where there is no gene transfer, then all genes share a common pattern of descent along the species tree. When genes can be transferred, they may share only part of their history with other gene families.

Gene tree-aware approaches were first used to deal with incomplete lineage sorting (ILS) through the multispecies coalescent \cite{Rannala2003a}. In that framework, all gene families have evolved within the boundaries of the species history, and heterogeneities among gene histories originate from population-level sorting of alleles only. More recently, similar models have been proposed to deal with other processes of genome evolution, namely gene duplication, transfer, and loss (DTL). For an in-depth review, please see \cite{Szollosi2014}. With these models, gene families can have a wider array of histories, and can differ drastically from the species tree. Invariably, whether they deal with ILS or DTL events, gene tree-species tree models have been found to produce gene trees that are more accurate than competing approaches. This is expected: as more information is used to reconstruct gene trees, stochastic error should diminish. 

Much like gene content approaches, gene tree-aware approaches can be based on probabilistic models that include parameters for DTL events \cite{Szollosi2012,Boussau2013a,Akerborg2009,Rasmussen2010}, or on parsimonious models, in which case DTL events are associated with costs \cite{Doyon2010,Bansal2014,Scornavacca2014}. Gene tree-species tree approaches however are computationally challenging. Interpreting a gene tree in the light of a species tree by placing events of gene duplication, transfer and loss, a process called \emph{reconciling} a gene tree, is not difficult provided rates or costs of events are provided. Things get more complicated when the gene tree is not assumed to be known, and needs to be reconstructed. Naturally, if the species tree itself also needs to be reconstructed, then the task becomes extremely difficult; however in the rest of this article we will assume the species tree is known without uncertainty. 

Methods to reconstruct gene trees using gene tree-aware approaches can use tree exploration heuristics similar to those found in commonly-used programs for phylogenetic tree reconstruction \cite{Guindon2010a,Stamatakis2005,Ronquist2012,Drummond2006,Lartillot2009}, as in Phyldog \cite{Boussau2013a} or in DLRS \cite{Sjostrand2012}. These approaches however tend to be slow, which motivated other approaches based on the consideration of a set of candidate gene trees obtained using faster approaches that do not consider a species tree. These approaches include TreefixDTL \cite{Bansal2014}, ALE \cite{Szollosi2013,Szollosi2013a} and  TERA \cite{Scornavacca2014}. The latter two approaches are extensions of an idea initially proposed in \cite{David2010} and formalized in \cite{Szollosi2013} and are particularly fast and accurate. They are based on the "amalgamation" idea. Based on a sample of gene trees, amalgamation is a dynamic programming algorithm that allows the exhaustive exploration of  a large space of gene trees. In fact, based on a limited set of gene trees, amalgamation allows considering a much larger space of gene trees, because it can piece together clades from several trees at a time to generate new trees, not present in the initial sample of gene trees. This technique is found to improve on competing approaches \cite{Szollosi2013,Scornavacca2014} in both speed and accuracy. 

Probabilistic gene tree-aware approaches can also be used to date trees. In such cases, gene tree-aware models often reconstruct ultrametric gene trees, a model describing the rate of sequence evolution needs to be used, and an ultrametric species tree whose nodes are anchored in time is required \cite{Rannala2003a,Heled2010,L2007,Arvestad2003a,Rasmussen2010}. Although these models contain additional parameters that need to be estimated, and are therefore computationally more complex to handle, they provide the ability to date events of gene family evolution along with the ability to estimate rates of events. Rates of events can then be compared across clades, although Fig.\ \ref{fig:2Approaches} is here to remind us that taxonomic sampling can have a non-trivial impact on the rates of reconstructed events. Another set of approaches avoids modelling the rate of sequence evolution and yet anchors events in time \cite{Szollosi2012,Doyon2010}. These approaches use a rooted ultrametric species tree in which nodes are ordered relative to each other, and mandate that transfers occur only between contemporaneous lineages. Gene trees however do not need to be ultrametric, which makes it possible to avoid using a model describing the rate of sequence evolution. Whether they use models describing the rate of sequence evolution or not, models that use ultrametric species trees are more realistic than models in which the nodes of the species tree are not ordered, because they include the constraint that only contemporaneous lineages can exchange genes; however, this constraint comes at a high computational cost.

\subsubsection{The impact of incomplete taxonomic sampling}
No matter how complex our models of genome evolution, our inferences depend on the sampling of our data set (Fig.\ \ref{fig:2Approaches}). Although progress in sequencing methods is moving at a fast pace, and genome sequences keep accumulating in databases, we will always be missing a clade or species that will prevent our data sets from being complete. It is unclear how such missing data impacts our inferences. Fig.\ \ref{fig:2Approaches} shows that missing species can lead to incorrectly interpret transfer events as duplication events, both for gene content and gene tree-aware approaches, but the magnitude of this effect is unknown. Worse, if our sampling of a clade misses a group of species with idiosyncratic characteristics (e.g. larger genomes, larger rates of gene transfer...), then our estimate of the parameters of genome evolution for this group will be biased. In the hope of achieving an unbiased estimate of genome evolution, it is important to try to quantify the bias imposed by incomplete taxonomic sampling.

%\subsubsection{Genome-wide reconstruction of gene histories to study genome evolution}
%Genome-wide applications of gene tree-species tree approaches or of gene content approaches allows uncovering general patterns of gene duplication, transfer and loss that affected all the gene families. This allows detecting branches that for instance show particularly high numbers of gene duplications, gene losses, or of gene transfers. There are two reasons why these numbers of events could be particularly high or low on a particular branch. First, this branch could be particularly long or short, respectively, meaning that a large or small amount of time may have passed between two speciations. Second, this branch could have high rates of duplication, transfer or loss, in which case it becomes interesting to investigate what biological parameters may be linked to these high rates.

\subsubsection{Comparing gene tree-aware and unaware approaches}
Although reconstructing genome evolution is a widely pursued endeavour, there have been few assessments of the inference methods used to reconstruct gene histories along the species tree. In this article we compare gene-content approaches with gene tree-aware approaches by using publicly-available software on two well-known clades in the tree of life. We use a state-of-the-art probabilistic gene-content approach, \COUNT~ \cite{Csuros2006,Csuros2010}, and a probabilistic gene tree-aware approach, \ALE~ (available at https://github.com/ssolo/ALE), adapted to handle undated species trees. We address the impact of incomplete taxonomic sampling by performing rarefaction studies, whereby species are pruned from our species trees and DTL rates are compared across samples. Our primary aim is to focus on the inferences of the two methods and explain their differences in the light of their strengths and shortcomings. In the process, we will contrast genome evolution in Cyanobacteria and Fungi.

\subsection{Genome evolution in Fungi and Cyanobacteria}

Fungi and Cyanobacteria \textit{a priori} differ in the way their genomes have evolved. For instance Fungi undergo whole genome duplications, whereas such events have not been reported in Cyanobacteria. While gene transfer has been claimed to occur in both Cyanobacteria and Fungi, it is unclear how frequent this process has been in these two clades. Another question of interest concerns highways of gene transfers, \textit{i.e.} pairs of branches or clades that appear to have undergone a high amount of gene transfers. While several highways of gene transfers have been claimed to exist in Bacteria, including in Cyanobacteria \cite{RG2005}, it is unknown whether there are highways of gene transfers in Fungi as well.

Both Cyanobacteria and Fungi have been the focus of several studies addressing genome evolution, because they display a wide variety in cell types and genome sizes, and because they have had an important environmental impact throughout their history. In the context of this article, these clades constitute excellent case studies to assess the behaviour of gene content and gene tree-aware approaches because of their wide diversity in genome size, along with the fact that different evolutionary dynamics are expected in Eukaryotes and Bacteria.

\subsubsection{Genome evolution in Fungi}
Fungi are characterized by two life forms: one, yeast-like, is unicellular. The other is multicellular and includes fungi with macroscopic fruiting bodies as well as filamentous fungi. In this study, we focus on the clade Dikarya, a subkingdom of fungi that account for roughly 98\% of described species. This clade is composed of two well-characterized phyla, Basidiomycota and Ascomycota. We  use the genome sequences included in the HOGENOM database \cite{Penel2009}. These two clades display a wide variety in genome sizes (from 5,200 to 10,000 protein coding genes, approximately), and have a phylogeny that can be unambiguously rooted between Basidiomycota and Ascomycota. Studies of genome evolution in these clades have focused for instance on the impact of whole genome duplications \cite{Wapinski2007}, on the evolution of the yeast (unicellular) form \cite{Nagy2014}, or on the evolution of pathways for the decomposition of plant material \cite{Eastwood2011,Floudas2012}. Recently, there have been reports of noticeable amounts of gene transfers in Fungi \cite{Slot2011,Richards2011a,Richards2011}. In particular several examples indicate that the \emph{Aspergillus} genome has been "sculpted by gene transfer"\cite{Gibbons2013}. This is consistent with reports that lateral gene transfers have been important throughout eukaryotic evolution \cite{Hirt2015}.

\subsubsection{Genome evolution in Cyanobacteria}
Cyanobacteria contain both unicellular organisms as well as organisms with two cell types, or that organize in filaments, which makes them unique among Prokaryotes for their ability to leave a recognizable trace in the fossil record \cite{Tomitani2006}. They display a wide range in genome size (from 1,200 to 4,500 protein coding genes, approximately), and have had a lasting impact on the Earth with the release of massive amounts of oxygen in the atmosphere billions of years ago \cite{Kaufman2007}. From a phylogenomics perspective cyanobacterial genomes share a relatively large core genome that allows the reconstruction of a well supported species phylogeny despite the antiquity of the phylum. Cyanobacteria have also served as a model system for investigating horizontal gene transfer \cite{Szollosi2012}, and have been reported to display highways of gene transfers \cite{RG2005}.    

\section{Methods}

\subsection{Data set construction}
\subsubsection{Fungi}
First, we selected all the species belonging to Fungi in the HOGENOM6 database\cite{Penel2009}, yielding 32  species. We retrieved the protein sequences clustered into homologous gene families (21701 families, discarding the very large families HOG100000000, HOG200000000 or HOG300000000, for which no alignment is available in the database). We discarded 8662 families containing only 2 or 1 genes  from Fungi. Gene trees were constructed for 1791 families containing only three genes (triplets), for which a single topology is possible. We aligned all families with 4 sequences or more using MUSCLE \cite{Edgar2004} with default parameters and selected reliably aligned sites using GBLOCKS \cite{Castresana2000}.  The parameters employed were "minimum number of sequences for a conserved position" b1 = 50, "minimum number of sequences for a flank position" b2 = 50 and "allowed gap positions" b5 = a (all). To estimate computing time per family, we measured the time PhyloBayes took \cite{Lartillot2009} to compute 10 trees based on each alignment. We discarded the decile of the slowest families. For each remaining alignment we ran 2 chains using PhyloBayes, calculating 5500 gene trees (discarding  the first 500 as burn-in), using the LG model of evolution \cite{SQ2008}. In the end we were able to compute at least one chain for  9596 gene families. Combined with the 1791 triplets, our data set contains in total 11387 gene families, totaling 135346 genes, while 24327 genes were discarded during our selection process (not counting the three HOGENOM families without alignments). 

Given that the tree of Fungi is still unresolved with Microsporidia branching in an undefined place,  we decided to use a smaller dataset, comprising the clade of Dikarya (28 species). This has the advantage that this clade can be easily rooted between Ascomycota and Basidiomycota. We pruned the gene trees removing from them two species of Microsporidia as well as \textit{Allomyces macrogynus} and \textit{Spizellomyces punctatus}, which belong to other basal clades of Fungi. In total, we used 11295 gene families. The gene trees are well resolved with an average posterior support of 0.97 (median=1).

Due to the uncertain position of \textit{Aspergillus nidulans}, we relied on two species trees: one reconstructed from a concatenate, and one drawn from the literature. For our first tree, which we call tree A, we used a concatenate of 529 near universal single-copy gene family alignments (25 or more species represented out of 28). In total the alignment contained 221 127 amino-acid sites including 24 514 without missing data. Both PhyML \cite{Guindon2010a} using the LG model of evolution \cite{SQ2008} and a Gamma distribution to account for rate variation \cite{Yang1994} and Phylobayes \cite{Lartillot2009} using the CAT model with Poisson exchangeabilities \cite{N2004} recovered the same topology. We rooted the species tree between Ascomycota and Basidiomycota. The resulting phylogeny identifies the major clades, Pezizomycotina, which groups \textit{Neurospora crassa} and \textit{Aspergillus} fungi, and Saccharomycotina, which notably groups \textit{Yarrowia lipolytica}, \textit{Candida} species, and \textit{Saccharomyces}. Our phylogeny is in agreement with the phylogenies of \cite{Fitzpatrick2006}, but in their study the position of \textit{Aspergillus nidulans} changes depending on the method: a concatenate based on 153 universal genes places it next to \textit{Aspergillus fumigatus}, as we do, but supertree methods find it at the base of the \textit{Aspergillus} clade. To account for these discrepancies, we reconstructed a second species tree where \textit{Aspergillus nidulans} is at the base of the \textit{Aspergillus} clade. We call this tree B, and estimated its branch lengths using PhyML with the same model as above. Tree A and Tree B can be found in the supplementary material.

\subsubsection{Cyanobacteria}
We selected all the species belonging to Cyanobacteria in the HOGENOM6 database\cite{Penel2009}, yielding 40  species. We reconstructed an unrooted species phylogeny using PhyML \cite{Guindon2010a} using the LG model of evolution \cite{SQ2008} and a gamma distribution to account for rate variation \cite{Yang1994} from a concatenate of 470 near universal single-copy gene family alignments (38 or more species represented out of 40). In total the alignment contained 126 180 amino-acid sites including 67 646 without missing data. The resulting tree agreed with our genome-scale reconstruction \cite{Szollosi2012} and other previous phylogenomic results (see discussion in \cite{Szollosi2012}, available in the supplementary material). We rooted the species tree according to \cite{Szollosi2012}. For 7415 gene families with 3 or more genes we employed the alignment procedure and sampling procedure described above. The gene trees are well resolved with an average posterior support of 0.96 (median=1).

\subsection{Inference methods}

\subsubsection{\COUNT}

\COUNT\ 
is a software package for performing studies in gene family evolution. It can perform ancestral genome reconstruction by posterior probabilities in a phylogenetic birth-and-death model \cite{Csuros2010}. Rates were optimized using a Gain-loss-duplication model, with default parameters and allowing different gain-loss and duplication-loss rates for different branches. One hundred rounds of optimization were computed. 
%The same families that we employed for the computation of the ALEml model were used.   

\subsubsection{\ALE}

\ALE\
implements a probabilistic approach to exhaustively explore all reconciled gene trees that can be amalgamated as a combination of clades observed in a sample of gene trees \cite{Szollosi2013a} in the context of different species tree-gene tree reconciliation models, in particular the model described in \cite{Szollosi2013}, which allows for the duplication, transfer and loss of genes. ALE can be used to efficiently approximate the sum of the joint likelihood over amalgamations and to find the reconciled gene tree that maximizes the joint likelihood among all such trees or sample the space of possible reconciliations. Here, we use two reconciliation methods, a simplified Duplication, Transfer and Loss (DTL) approach that does not consider the temporal information from the species tree, and a version of this model that only allows Duplication and Loss (DL). These methods are available as part of the open-source ALE project ( https://github.com/ssolo/ALE ).

\subsection{Analyses}

Highways were identified between pairs of species that exchange large numbers of genes. The number of genes exchanged was averaged over 100 reconciliations drawn from \ALE using the program ALEsample, and summed across all gene families in our datasets. 

Synteny information was extracted from gene positions in the genomes. Pairwise comparisons of genomes between species were performed. Synteny was found to be conserved if a gene had as a neighbor a gene whose ortholog was also its own ortholog's neighbor. For simplicity, only gene families with one gene per species were considered in the synteny analyses. For a given pairwise comparison, a gene was declared as non-transferred if, along the path between two species, no transfer event had affected the gene of each species, and declared as transferred otherwise.

\section{Results}

\subsection{General patterns of genome evolution}
\begin{figure}
\centering

\includegraphics[width=0.9\textwidth]{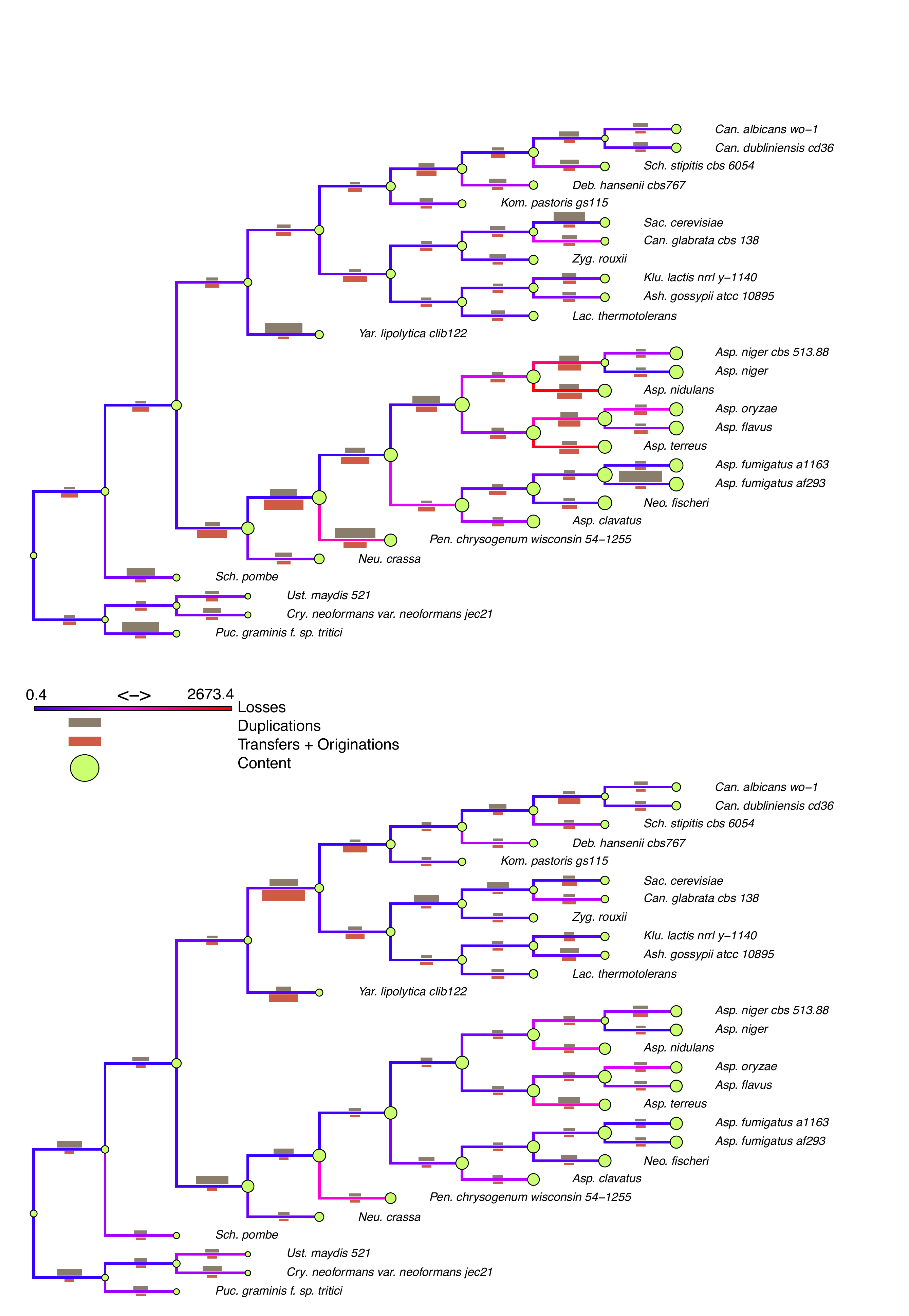}

\caption{\label{fig:FungiDTLOri}Genome evolution in Ascomycota and Basidiomycota (tree A). Edges are color-coded according to the inferred numbers of losses along the branches. Crimson bars represent numbers of gene gains (transfers + originations) arriving on the branch; taupe bars represent numbers of duplications happening on the branch. At each node, genome content size is represented as a green disk. Top: inferences from \ALE. Bottom: inferences from \COUNT. The corresponding graph for tree B is available in the supplementary material.}
\end{figure}

\begin{figure}
\centering
\begin{subfigure}[b]{1\textwidth}
\includegraphics[width=0.9\textwidth]{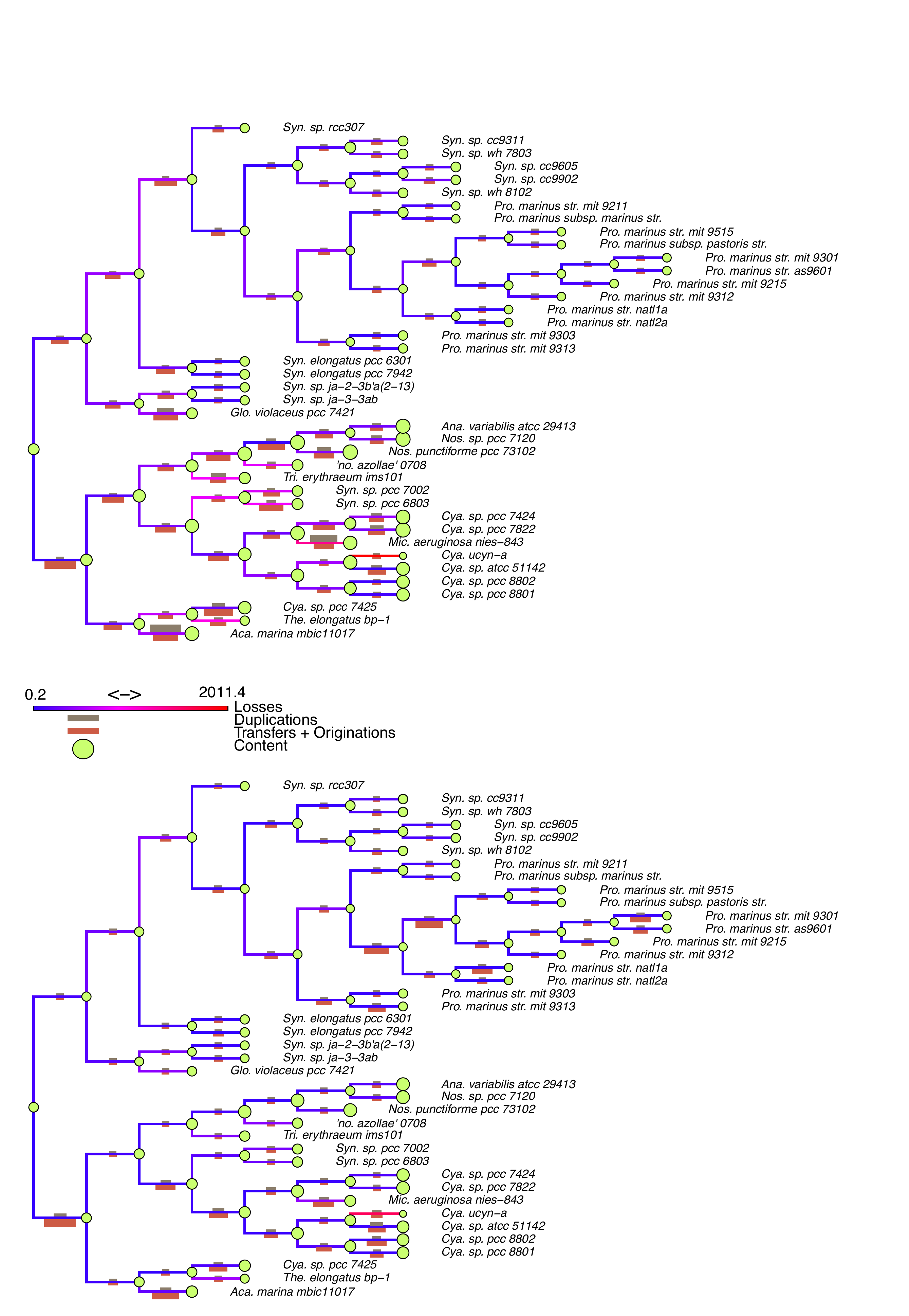}
\end{subfigure}
\caption{\label{fig:CyanoDTLOri}Genome evolution in Cyanobacteria. Edges are color-coded according to the inferred numbers of losses along the branches. Crimson bars represent numbers of gene gains (transfers + originations) arriving on the branch; taupe bars represent numbers of duplications happening on the branch. At each node, genome content size is represented as a green disk. Top: inferences from \ALE. Bottom: inferences from \COUNT.}
\end{figure}

Fig.\ \ref{fig:FungiDTLOri} and \ref{fig:CyanoDTLOri} show the reconstruction of genome evolution across Fungi and Cyanobacteria, respectively, using both \COUNT ~and \ALE. Although \COUNT~ and \ALE~ differ in their input data and in the types of events they can detect, their inferences are qualitatively  similar, finding comparable genome size dynamics, and proportions of events on branches. 

In both Cyanobacteria and Fungi, with both methods, a clade with large genomes (multicellular \textit{Aspergillus} clade of molds and  the clade including freshwater and multicellular cyanobacteria such as \emph{Nostoc} and \emph{Cyanothece}) and a clade with smaller genomes (unicellular clade of yeasts including \emph{Saccharomyces} and \emph{Candida} and unicellular planktonic cyanobacteria including \emph{Prochlorococcus} and \emph{Synechococcus}) can be observed.  

In Fungi, the clade with large genomes (fungi from the multicellular \textit{Aspergillus} clade of molds) shows a large portion of gene transfers on several of its branches, whereas gene transfers appear much less prevalent in the clade with small genomes (containing the unicellular yeasts). This result confirms earlier reports based on smaller data sets of larger amounts of gene transfers in the \textit{Aspergillus} clade than in the yeast clade \cite{Richards2011}. Several branches show an excess of gene duplications compared to gene transfers. Although a whole genome duplication (WGD) occurred in the ancestor of \textit{Saccharomyces cerevisiae} and \textit{Candida glabrata} \cite{Wapinski2007}, both models fail to pick an increased amount of gene duplications on the relevant branch. They recover an increased amount of duplications on the branch leading to \textit{Saccharomyces cerevisiae} alone, possibly because \textit{Candida glabrata} has lost a large number of genes, which, in the absence of synteny (which was used by \cite{Wapinski2007} to detect the WGD), may have erased a large part of the signal supporting the whole genome duplication. The ancestor of all Dikarya is predicted to have a very small genome, which is likely the consequence of our unbalanced taxonomic sampling, with only 3 Ascomycota. Due to this design, families present only in 2 Ascomycota have been discarded from our data set, and therefore cannot be inferred at the root.

In Cyanobacteria, both the clades with large and small genomes appear to have similar genome dynamics, with more gene transfers than gene duplications. The ancestor of Cyanobacteria is predicted to have intermediate genome content in between that of the clade with small genomes (containing \textit{Prochlorococcus} species) and that of the clade with larger genomes.

\subsection{Gene tree-aware approaches are more sensitive than gene content approaches}

\begin{figure}
\centering
\includegraphics[width=1\textwidth]{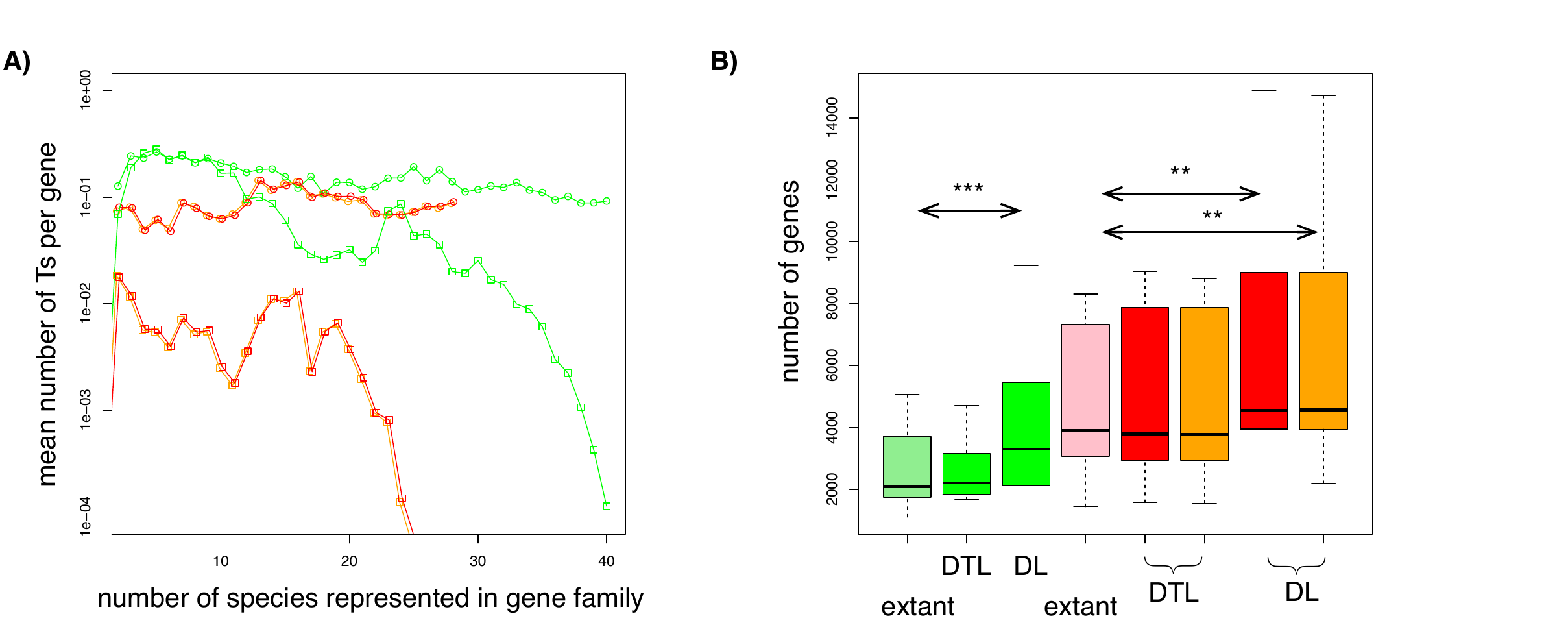}
\caption{\label{fig:fig5} {\bf Comparison between alternative genome evolution reconstruction methods.} (A) Comparing methods that consider transfers alongside duplications and losses, we find that the gene-tree aware method (\ALE, circles) infers an order of magnitude more transfers for Fungi. The gene-tree unaware method (\COUNT, squares) loses its ability to detect transfers as gene families become universal or near universal (28 species for Fungi, 40 for Cyanobacteria), while the gene-tree aware method does not. (B) To contrast gene-tree aware reconstruction that considers transfer (DTL) to one that only accounts for duplications and losses, but does not consider transfer (DL) we compare the number of genes in ancestral genomes to the number of genes in extant genomes. The DTL method infers ancestral genome sizes that fall within the distribution for extant genomes ($p>0.5$ two-sided Wilcox test) for both Fungi and Cyanobacteria. In contrast gene numbers based on the DL method are systematically  larger than extant ones ($p<10^{-3}$ for cyanobacteria and $p=0.02$ for Fungi with a one-sided Wilcox test for both tree A and B). Fungi are in dark grey (tree A red, tree B orange online) and Cyanobacteria in light grey (green online) throughout. }
\end{figure}

By design, gene tree-aware approaches can detect more events than gene-content approaches (Fig.\ \ref{fig:2Approaches}). Consistently, \ALE\ finds significantly more transfers than \COUNT, with \ALE~ finding an average of $0.16$ and $0.07$ transfers per gene in respectively Cyanobacteria and Fungi, in contrast to \COUNT, which finds $0.14$ and $0.06$.  It is difficult to determine how many of the additional transfers are due to \ALE\ finding true transfer events that \COUNT~ failed to detect and how many result from errors in reconstructed gene trees. Simulations do indicate that ALE recovers an unbiased estimate of the number of transfers \cite{Szollosi2013}, and in the case of Cyanobacteria reduces the number of inferred transfers by approximately two-thirds compared to gene trees reconstructed without the species tree (by PhyML\cite{Guindon2010a}).  Furthermore, Fig.\ \ref{fig:fig5}A shows that for cyanobacterial famillies represented in $8$ or fewer genomes the \ALE\ 
and \COUNT\ 
estimates closely agree. Regardless of potential overestimation of the number of transfers, Fig.\ \ref{fig:fig5}A also highlights that the number of transfer events per gene family is more-or-less homogeneous with respect to the number of species represented in the gene family for \ALE, but systematically decreases for \COUNT\ as more complete taxonomic distributions are approached. We see no biological reason why e.g.\ families with a complete taxonomic distribution would undergo much fewer transfers compared to families with slightly incomplete taxonomic sampling. Instead, we believe this effect results from a shortcoming of gene tree unaware methods, such as  \COUNT, whereby they are not able to infer transfer among families with complete taxonomic distribution, and progressively lose signal as complete taxonomic distribution is approached. 

\subsection{Duplication and loss methods systematically overestimate ancestral gene content}

The effect of gene transfers on gene phylogenies can be mimicked by a combination of gene duplications and losses. Therefore gene duplications and losses may be sufficient to account for genome dynamics in our two clades, and it is legitimate to ask about the need to incorporate transfers.  As shown in Fig.\ \ref{fig:fig5}B comparison of the gene content of extant genomes and reconstructions based on gene-tree aware reconstruction that considers transfer (DTL) shows that these methods reconstruct ancestral gene contents that are similar to those observed for extant genomes. In stark contrast, gene-tree aware reconstructions that only account for duplications and losses, but do not consider transfer (DL), infer systematically more genes to have been present in ancestral genomes than in extant ones. The largest ancestral genomes are inferred by DL methods in the common ancestors of clades where DTL methods predict the most transfers: the deepest node in the \emph{Aspergillus} genus has ancestral gene contents of  14244 genes according to the DL estimate, in comparison to 8238 genes for the DTL based estimate. The extant gene content in the genus in our sample is between 7797 (Aspergillus fumigatus A1163) and 8891 (\emph{Aspergillus terreus}).

\subsection{Rates of transfers are similar in Fungi and Cyanobacteria}

\begin{figure}
\centering
\includegraphics[width=1.\textwidth]{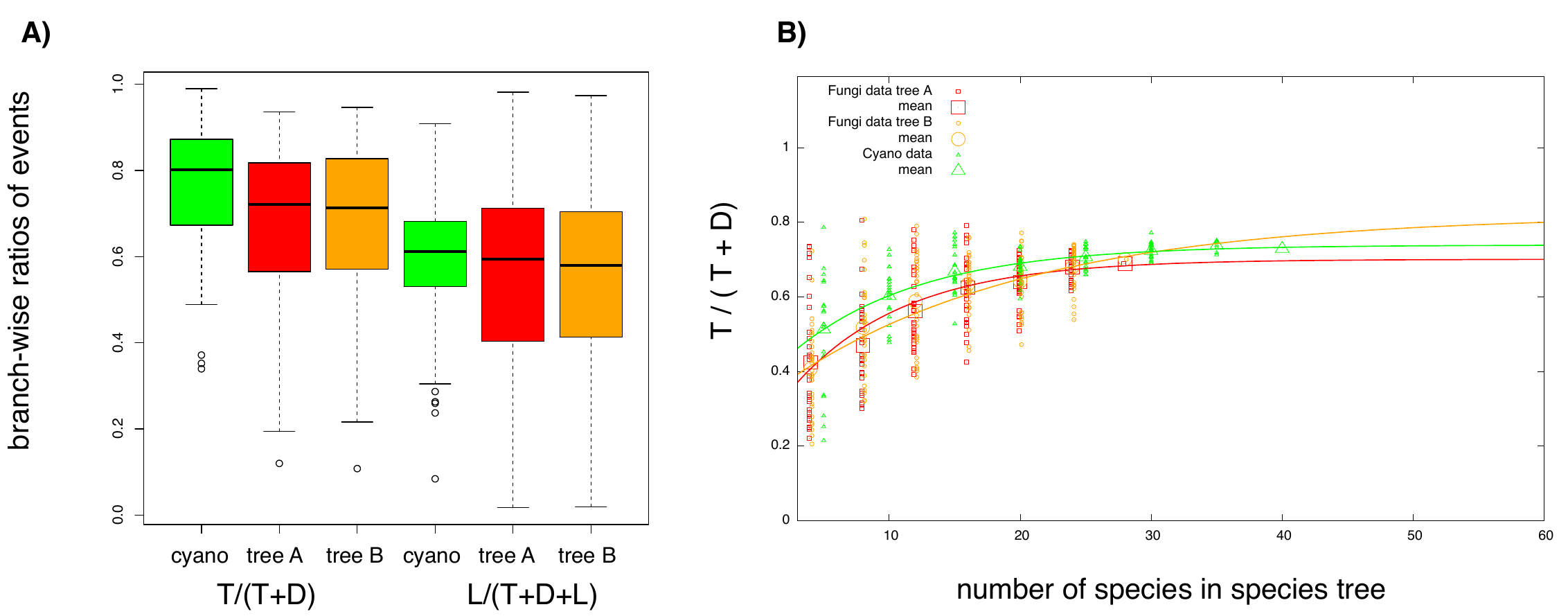}
\caption{\label{fig:fig4} {\bf Similar rate of transfers in Fungi and Cyanobacteria.} The cyanobacterial and fungal datasets we considered differ both in the number of genomes (40 vs. 28) and in their age (nearly 3 billion years vs. less than three quarters of a billion years). (A) In order to account for differences in age we consider the ratio $T/(T+D)$, \textit{i.e.} the fraction of transfer events among all gene birth events (duplications plus transfers) and $L/(T+D+L)$ the ratio of loss events to all events. (B) To ascertain the effect of the differences in the number of genomes we constructed replicates with random subsamples of genomes of varying numbers. Extrapolation of the results suggests that $T/((D+T)$ for Fungi is as large or larger than for Cyanobacteria (for details see text). Fungi are shown with squares for tree A (red online), and circles for tree B (orange online) and Cyanobacteria with triangles (green online).}
\end{figure}

Rates of transfer in Fungi and in Cyanobacteria appear to be very similar, as shown by the \ALE\ 
inferences (Figure \ref{fig:fig4}A). This finding does not come from differences in the age of the clades as we compare ratios of numbers of events, which controls for age. It does not appear to come from incomplete sampling either, as Figure \ref{fig:fig4}B shows that predictions based on subsampling the species in each data set still converge to similar ratios of numbers of events for Fungi and Cyanobacteria. To extrapolate the $T/(T+D)$ values we fit an \emph{ad hoc} curve that reaches saturation exponentially starting from an initial value for $0$ species.  Using all subsampled replicates a least-squares Marquardt-Levenberg algorithm yielded the similar asymptotic values of $T/(T+D)$, with $0.8\pm 0.1$ (Fungi assuming tree A), $0.7\pm 0.03$ (Fungi assuming tree B) and $0.74\pm 0.01$ in Cyanobacteria. The same procedure for $L/(T+D+L)$ produced the slightly higher asymptotic value for Fungi of $0.582 \pm 0.01$ for tree A and  $0.595 \pm 0.01$ for tree B, compared to  $0.52 \pm 0.01$ compared to Cyanobacteria.

These genome-wide inferences confirm earlier reports based on manual analyses of smaller data sets that significant numbers of transfers occurred in Fungi, in particular in the \textit{Aspergillus} clade \cite{Slot2011,Richards2011a,Richards2011,Gibbons2013}. Overall, these data show that genomes in Prokaryotes and Eukaryotes are not undergoing fundamentally different dynamics. We consider that additional analyses of data sets for different clades of both Prokaryotes and Eukaryotes, using gene tree-aware approaches as in this work, would provide a more fine-grained, quantitative view of the dynamics of genome evolution across the entire tree of life.

\subsection{There are highways of gene transfer in Fungi}
\begin{figure}
\centering
\includegraphics[width=0.9\textwidth]{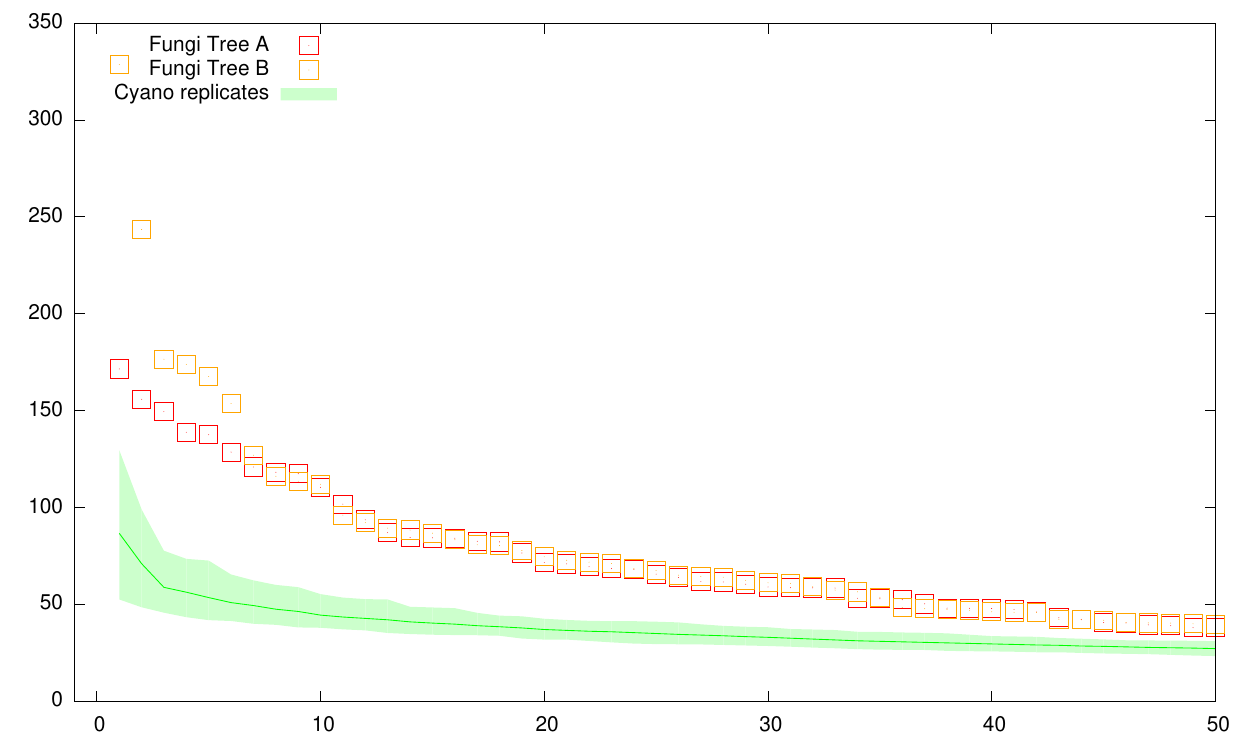}
\caption{\label{fig:fig6} {\bf Stronger highways in Fungi.}  Data points, dark gray (red online) for tree B and light gray (orange online) for tree A, correspond to numbers of transfers between pairs of branches  ("highways of transfers") in either Fungi phylogeny plotted in decreasing order. The continuous line (green online) shows the mean number of transfers between pairs of branches among 25 replicates where a random set of 28 cyanobacterial genomes were chosen as in Fig.\ \ref{fig:fig4}.B. The shaded area shows the 95\% confidence interval. Fungi are in dark grey (red online) and Cyanobacteria in light grey (green online) throughout. }
\end{figure}

\begin{figure}
\centering
\includegraphics[width=1\textwidth]{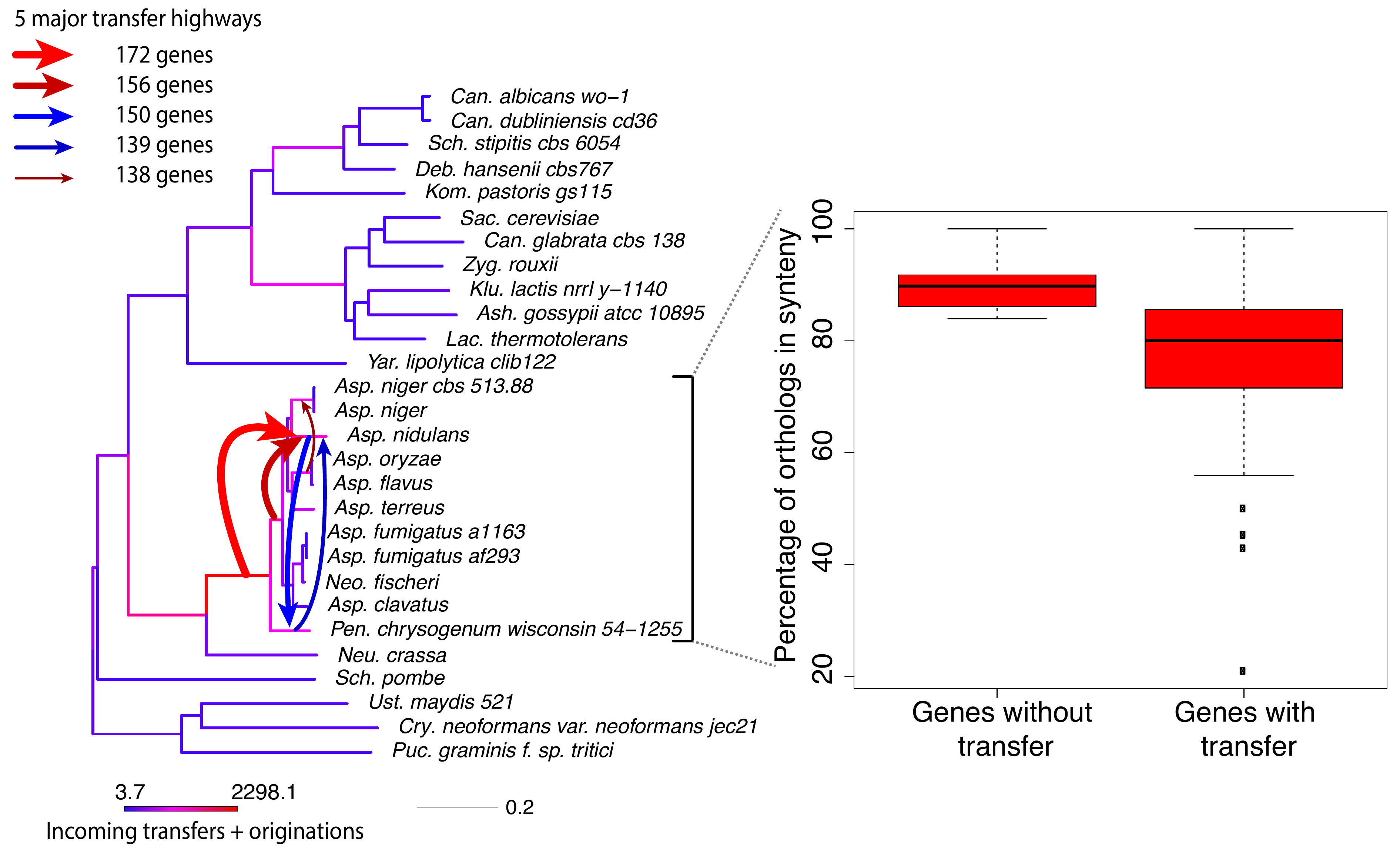}
\caption{\label{fig:fig7} {\bf Position of the 5 largest highways in Fungi on tree A.} Branch lengths are in expected numbers of amino-acid substitutions, and colored according to the number of incoming gene transfers and originations. The five largest highways of transfers are represented. On the right, boxplots show pairwise genome synteny comparisons. Transferred genes are found to change neighbors more often than non-transferred genes. On tree B, the five largest highways concern 329, 244, 177, 174, 168 genes, the three largest ones of which involve \textit{Aspergillus nidulans}.}
\end{figure}

One feature of genome evolution in Prokaryotes that has received considerable attention is the concept of highways of gene transfers \cite{RG2005,Bansal2013}. According to this model, some pairs of species or clades have exchanged large numbers of genes throughout their history, possibly because of a shared ecological niche. \ALE~ inferences provide us with an opportunity to look for such highways in Cyanobacteria and Fungi. Fig.\ \ref{fig:fig6} shows the distribution of the number of transfers per pairs of branches of the species tree in both Cyanobacteria and Fungi. Both distributions show a long tail, with many transfers occurring between branches that otherwise have exchanged little genetic material. However, some pairs of branches show very high numbers of gene transfer events. The heterogeneity is strongest in Fungi, where some pairs of branches are predicted to have undergone more than 150 gene transfers, or even 300 transfers on tree B (Fig. \ref{fig:fig7}). These transfers do not seem to be due to hybridization, as most of them are not replacement transfers, whereby a gene in a species is replaced by another gene coming from another species (the median branch-wise fraction of gene transfers that are compensated by loss on the same branch, \textit{i.e.} replacement transfers, is 31\% for Fungi on tree A , 34\% on tree B and 49\% for Cyanobacteria). For the same reason, these transfers cannot be misinterpreted events of incomplete lineage sorting. In fact, among genes that have only one ortholog per species, genes that have undergone a gene transfer tend to change position on the chromosome more often than genes that have not undergone a gene transfer (see Fig. \ref{fig:fig7}, right, for the \textit{Aspergillus} clade).  The pairs of branches with the largest numbers of transfers belong to the \textit{Aspergillus} clade, in agreement with the overall larger amount of transfers detected in this clade and in agreement with previous reports \cite{Gibbons2013}. The species involved in the largest number of transfers in either tree A or tree B is \textit{Aspergillus nidulans}, precisely the species whose position is contentious. This suggests that lateral gene transfers in Fungi may be significant enough to make  reconstruction of the species phylogeny difficult.  Although deeper sampling could change the numbers of gene transfers found on each pair of branches, for instance by breaking branches involved in a highway, it seems unlikely that the conclusion that there are branch pairs or group of branches exchanging large numbers of genes in Fungi would change.

\subsection{Genes tend to be transferred together}
The distribution of transferred genes along chromosomes appears to be consistent with the transfer of chromosomal segments that can include more than one gene. Counting only transfers to the terminal branches, transferred genes appear preferentially next to another transferred gene: on average, 4.7 times more often in Fungi (on tree A, 6.1 times more often on tree B), and 5.3 times more often in Cyanobacteria. Given that genes transferred on terminal branches make up a minority of the genomes, this means that transfers tend to affect blocks of several genes at a time.

\subsection{\ALE reconstructs accurate gene trees}
In \cite{Szollosi2013}, we found using realistic simulations that amalgamation of gene trees using a DTL model produced accurate gene trees: the number of duplications and transfers needed to reconcile our reconstructed trees was statistically indistinguishable from the corresponding number of events needed to reconcile the “real” trees that had been used to simulate gene alignments. In the present study, empirical results also show that gene trees reconstructed by \ALE are accurate. First, the fact that the reconstructions of ancestral genome sizes based on our reconciled gene trees are not significantly different from extant genome sizes suggests that our gene trees do not contain large numbers of incorrect bipartitions. Second, the overrepresentation of transferred genes in tandem cannot be explained by random errors in gene trees, but shows that \textit{bona fide} information can be retrieved from  gene trees reconstructed by \ALE.

\section{Conclusion}

Our genome-scale phylogenetic analysis of genome evolution in Cyanobacteria and Fungi shows that Fungi exhibit similar rates of transfers as Cyanobacteria, and display apparent highways of gene transfers. Whether these highways of gene transfers correspond to shared ecological niches or to particular mechanisms to incorporate foreign DNA remains to be investigated. In both clades, gene transfers appear to occur in blocks, not just one gene at a time. Further investigation of those transferred blocks of genes may prove useful for functional annotation, as co-transferred genes may be functionally related. 

This study also allows the comparative study of different methodologies for reconstructing genome evolution. We show that the recent developments provide a framework adapted to different domains of life, and that gene tree-aware methods show more precision in the quantification of gene transfers.

Our results suggest that further analyses of data sets for other clades of Prokaryotes and Eukaryotes, using gene tree-aware approaches , will provide a more fine-grained, quantitative view of the dynamics of genome evolution across the tree of life.

\section{Acknowledgments}
Data will be available on DataDryad upon acceptance. This project was supported by the French Agence Nationale de la Recherche (ANR) through Grant ANR-10-BINF-01-01 ``Ancestrome''.
 GJSz was supported by the FP7-PEOPLE-CIG grant "GENESTORY".

\bibliographystyle{prsb}
\bibliography{biblio}

\begin{thebibliography}{10}
\expandafter\ifx\csname urlstyle\endcsname\relax
  \providecommand{\doi}[1]{doi:\discretionary{}{}{}#1}\else
  \providecommand{\doi}{doi:\discretionary{}{}{}\begingroup
  \urlstyle{rm}\Url}\fi

\bibitem{Szollosi2014}
Szollosi, G.~J., Tannier, E., Daubin, V. \& Boussau, B., 2015 {The inference of
  gene trees with species trees}.
\newblock \emph{Systematic Biology} \textbf{64}, syu048----.
\newblock ISSN 1063-5157.
\newblock (\doi{10.1093/sysbio/syu048}).

\bibitem{Groussin2014}
Groussin, M., Hobbs, J.~K., Sz\"{o}llősi, G.~J., Gribaldo, S., Arcus, V.~L. \&
  Gouy, M., 2015 {Toward More Accurate Ancestral Protein Genotype-Phenotype
  Reconstructions with the Use of Species Tree-Aware Gene Trees}.
\newblock \emph{Molecular Biology and Evolution} \textbf{32}, 13--22.
\newblock ISSN 0737-4038.
\newblock (\doi{10.1093/molbev/msu305}).

\bibitem{Boussau2004}
Boussau, B., Karlberg, E.~O., Frank, A.~C., Legault, B.-A.~A. \& Andersson, S.
  G.~E., 2004 {Computational inference of scenarios for
  $\alpha$-proteobacterial genome evolution}.
\newblock \emph{Proceedings of the National Academy of Sciences of the United
  States of America} \textbf{101}, 9722.
\newblock ISSN 0027-8424.
\newblock (\doi{10.1073/pnas.0400975101}).

\bibitem{Dagan2007}
Dagan, T. \& Martin, W., 2007 {Ancestral genome sizes specify the minimum rate
  of lateral gene transfer during prokaryote evolution}.
\newblock \emph{Proceedings of the National Academy of Sciences} \textbf{104},
  870--875.
\newblock ISSN 1091-6490.

\bibitem{David2010}
David, L.~A. \& Alm, E.~J., 2010 {Rapid evolutionary innovation during an
  Archaean genetic expansion}.
\newblock \emph{Nature} \textbf{469}, 93--96.
\newblock ISSN 1476-4679.

\bibitem{O2010}
Cohen, O. \& Pupko, T., 2010 {Inference and characterization of horizontally
  transferred gene families using stochastic mapping}.
\newblock \emph{Molecular Biology and Evolution} \textbf{27}, 703--13.
\newblock ISSN 1537-1719.

\bibitem{Csuros2006}
Csűr\"{o}s, M. \& Mikl\'{o}s, I., 2006 {A probabilistic model for gene content
  evolution with duplication, loss, and horizontal transfer}.
\newblock \emph{Research in Computational Molecular Biology} \textbf{q-bio.PE},
  206--220.

\bibitem{Szollosi2012a}
Sz\"{o}llosi, G.~J. \& Daubin, V., 2012 {Modeling gene family evolution and
  reconciling phylogenetic discord}.
\newblock \emph{Methods in Molecular Biology} \textbf{856}, 29--51.
\newblock ISSN 10643745.
\newblock (\doi{10.1007/978-1-61779-585-5\_2}).

\bibitem{Rannala2003a}
Rannala, B. \& Yang, Z., 2003 {Bayes estimation of species divergence times and
  ancestral population sizes using DNA sequences from multiple loci}.
\newblock \emph{Genetics} \textbf{164}, 1645--1656.
\newblock ISSN 0016-6731.

\bibitem{Szollosi2012}
Sz\"{o}llosi, G.~J., Boussau, B., Abby, S.~S., Tannier, E. \& Daubin, V., 2012
  {Phylogenetic modeling of lateral gene transfer reconstructs the pattern and
  relative timing of speciations.}
\newblock \emph{Proceedings of the National Academy of Sciences of the United
  States of America} \textbf{109}, 17513--8.
\newblock ISSN 1091-6490.
\newblock (\doi{10.1073/pnas.1202997109}).

\bibitem{Boussau2013a}
Boussau, B., Sz\"{o}llosi, G.~J., Duret, L., Gouy, M., Tannier, E. \& Daubin,
  V., 2013 {Genome-scale coestimation of species and gene trees}.
\newblock \emph{Genome research} \textbf{23}, 323--330.
\newblock ISSN 1549-5469.
\newblock (\doi{10.1101/gr.141978.112}).

\bibitem{Akerborg2009}
Akerborg, O., Sennblad, B., Arvestad, L. \& Lagergren, J., 2009 {Simultaneous
  Bayesian gene tree reconstruction and reconciliation analysis}.
\newblock \emph{Proceedings of the National Academy of Sciences} \textbf{106},
  5714--5719.
\newblock ISSN 1091-6490.

\bibitem{Rasmussen2010}
Rasmussen, M.~D. \& Kellis, M., 2010 {A Bayesian Approach for Fast and Accurate
  Gene Tree Reconstruction}.
\newblock \emph{Molecular Biology and Evolution} \textbf{28}, 273--290.
\newblock ISSN 1537-1719.

\bibitem{Doyon2010}
Doyon, J.-p., Scornavacca, C., Gorbunov, K.~Y., Sz\"{o}llősi, G.~J., Ranwez,
  V. \& Berry, V., 2010 {An efficient algorithm for gene / species trees
  parsimonious reconciliation with losses, duplications, and transfers}.
\newblock In \emph{Comparative Genomics} (ed. E.~Tannier), volume 6398, pp.
  93--108. Springer Berlin Heidelberg.
\newblock ISBN 978-3-642-16180-3.
\newblock (\doi{10.1007/978-3-642-16181-0\_9}).

\bibitem{Bansal2014}
Bansal, M.~S., Wu, Y.-C., Alm, E.~J. \& Kellis, M., 2014 {Improved gene tree
  error correction in the presence of horizontal gene transfer}.
\newblock \emph{Bioinformatics} pp. 1--8.
\newblock ISSN 1367-4803.
\newblock (\doi{10.1093/bioinformatics/btu806}).

\bibitem{Scornavacca2014}
Scornavacca, C., Jacox, E. \& Sz\"{o}llősi, G.~J., 2015 {Joint amalgamation of
  most parsimonious reconciled gene trees}.
\newblock \emph{Bioinformatics} \textbf{31}, 841--848.
\newblock ISSN 1367-4803.
\newblock (\doi{10.1093/bioinformatics/btu728}).

\bibitem{Guindon2010a}
Guindon, S., Dufayard, J.-F., Lefort, V., Anisimova, M., Hordijk, W. \&
  Gascuel, O., 2010 {New algorithms and methods to estimate maximum-likelihood
  phylogenies: assessing the performance of PhyML 3.0.}
\newblock \emph{Systematic Biology} \textbf{59}, 307--321.
\newblock ISSN 1076-836X.

\bibitem{Stamatakis2005}
Stamatakis, A., Ludwig, T. \& Meier, H., 2005 {RAxML-III: a fast program for
  maximum likelihood-based inference of large phylogenetic trees}.
\newblock \emph{Bioinformatics} ISSN 1460-2059.

\bibitem{Ronquist2012}
Ronquist, F., Teslenko, M., van~der Mark, P., Ayres, D.~L., Darling, A.,
  H\"{o}hna, S., Larget, B., Liu, L., Suchard, M.~A. \& Huelsenbeck, J.~P.,
  2012 {MrBayes 3.2: Efficient Bayesian Phylogenetic Inference and Model Choice
  Across a Large Model Space.}
\newblock \emph{Systematic Biology} \textbf{61}, 539--542.

\bibitem{Drummond2006}
Drummond, A.~J., Ho, S. Y.~W., Phillips, M.~J. \& Rambaut, A., 2006 {Relaxed
  phylogenetics and dating with confidence}.
\newblock \emph{PLoS Biology} \textbf{4}, e88.
\newblock ISSN 1545-7885.

\bibitem{Lartillot2009}
Lartillot, N., Lepage, T. \& Blanquart, S., 2009 {PhyloBayes 3: a Bayesian
  software package for phylogenetic reconstruction and molecular dating}.
\newblock \emph{Bioinformatics} \textbf{25}, 2286.
\newblock ISSN 1460-2059.

\bibitem{Sjostrand2012}
Sj\"{o}strand, J., Sennblad, B., Arvestad, L. \& Lagergren, J., 2012 {DLRS:
  gene tree evolution in light of a species tree.}
\newblock \emph{Bioinformatics (Oxford, England)} \textbf{28}, 2994--5.
\newblock ISSN 1367-4811.
\newblock (\doi{10.1093/bioinformatics/bts548}).

\bibitem{Szollosi2013}
Sz\"{o}llosi, G.~J., Rosikiewicz, W., Boussau, B., Tannier, E. \& Daubin, V.,
  2013 {Efficient exploration of the space of reconciled gene trees}.
\newblock \emph{Systematic Biology} \textbf{62}, 901--912.
\newblock ISSN 10635157.
\newblock (\doi{10.1093/sysbio/syt054}).

\bibitem{Szollosi2013a}
Sz\"{o}llosi, G.~J., Tannier, E., Lartillot, N. \& Daubin, V., 2013 {Lateral
  gene transfer from the dead}.
\newblock \emph{Systematic Biology} \textbf{62}, 386--397.
\newblock ISSN 10635157.
\newblock (\doi{10.1093/sysbio/syt003}).

\bibitem{Heled2010}
Heled, J. \& Drummond, A.~J., 2010 {Bayesian inference of species trees from
  multilocus data.}
\newblock \emph{Molecular Biology and Evolution} \textbf{27}, 570--80.
\newblock ISSN 1537-1719.
\newblock (\doi{10.1093/molbev/msp274}).

\bibitem{L2007}
Liu, L. \& Pearl, D.~K., 2007 {Species trees from gene trees: reconstructing
  Bayesian posterior distributions of a species phylogeny using estimated gene
  tree distributions.}
\newblock \emph{Systematic biology} \textbf{56}, 504--514.
\newblock ISSN 1063-5157.
\newblock (\doi{10.1080/10635150701429982}).

\bibitem{Arvestad2003a}
Arvestad, L., Berglund, A.-C., Lagergren, J. \& Sennblad, B., 2003 {Bayesian
  gene/species tree reconciliation and orthology analysis using MCMC}.
\newblock \emph{Bioinformatics} \textbf{19}, i7--i15.
\newblock ISSN 1367-4803.
\newblock (\doi{10.1093/bioinformatics/btg1000}).

\bibitem{Csuros2010}
Csur\"{o}s, M., 2010 {Count: Evolutionary analysis of phylogenetic profiles
  with parsimony and likelihood}.
\newblock \emph{Bioinformatics} \textbf{26}, 1910--1912.
\newblock ISSN 13674803.
\newblock (\doi{10.1093/bioinformatics/btq315}).

\bibitem{RG2005}
Beiko, R.~G., Harlow, T.~J. \& Ragan, M.~A., 2005 {Highways of gene sharing in
  prokaryotes.}
\newblock \emph{Proceedings of the National Academy of Sciences of the United
  States of America} \textbf{102}, 14332--14337.
\newblock ISSN 0027-8424.
\newblock (\doi{10.1073/pnas.0504068102}).

\bibitem{Penel2009}
Penel, S., Arigon, A.-M., Dufayard, J.-F., Sertier, A.-S., Daubin, V., Duret,
  L., Gouy, M. \& Perri\`{e}re, G., 2009 {Databases of homologous gene families
  for comparative genomics.}
\newblock \emph{BMC Bioinformatics} \textbf{10 Suppl 6}, S3.
\newblock ISSN 14712105.

\bibitem{Wapinski2007}
Wapinski, I., Pfeffer, A., Friedman, N. \& Regev, A., 2007 {Natural history and
  evolutionary principles of gene duplication in fungi}.
\newblock \emph{Nature} \textbf{449}, 54--61.
\newblock ISSN 1476-4679.

\bibitem{Nagy2014}
Nagy, L.~G., Ohm, R.~a., Kov\'{a}cs, G.~M., Floudas, D., Riley, R., G\'{a}cser,
  A., Sipiczki, M., Davis, J.~M., Doty, S.~L., de~Hoog, G.~S. \emph{et~al.},
  2014 {Latent homology and convergent regulatory evolution underlies the
  repeated emergence of yeasts.}
\newblock \emph{Nature communications} \textbf{5}, 4471.
\newblock ISSN 2041-1723.
\newblock (\doi{10.1038/ncomms5471}).

\bibitem{Eastwood2011}
Eastwood, D.~C., Floudas, D., Binder, M., Majcherczyk, A., Schneider, P.,
  Aerts, A., Asiegbu, F.~O., Baker, S.~E., Barry, K., Bendiksby, M.
  \emph{et~al.}, 2011 {The plant cell wall-decomposing machinery underlies the
  functional diversity of forest fungi.}
\newblock \emph{Science (New York, N.Y.)} \textbf{333}, 762--5.
\newblock ISSN 1095-9203.
\newblock (\doi{10.1126/science.1205411}).

\bibitem{Floudas2012}
Floudas, D., Binder, M., Riley, R., Barry, K., Blanchette, R.~A., Henrissat,
  B., Mart\'{\i}nez, A.~T., Otillar, R., Spatafora, J.~W., Yadav, J.~S.
  \emph{et~al.}, 2012 {The Paleozoic origin of enzymatic lignin decomposition
  reconstructed from 31 fungal genomes.}
\newblock \emph{Science (New York, N.Y.)} \textbf{336}, 1715--9.
\newblock ISSN 1095-9203.
\newblock (\doi{10.1126/science.1221748}).

\bibitem{Slot2011}
Slot, J.~C. \& Rokas, A., 2011 {Horizontal transfer of a large and highly toxic
  secondary metabolic gene cluster between fungi}.
\newblock \emph{Current Biology} \textbf{21}, 134--139.
\newblock ISSN 09609822.
\newblock (\doi{10.1016/j.cub.2010.12.020}).

\bibitem{Richards2011a}
Richards, T.~a., Leonard, G., Soanes, D.~M. \& Talbot, N.~J., 2011 {Gene
  transfer into the fungi}.
\newblock \emph{Fungal Biology Reviews} \textbf{25}, 98--110.
\newblock ISSN 17494613.
\newblock (\doi{10.1016/j.fbr.2011.04.003}).

\bibitem{Richards2011}
Richards, T.~a., 2011 {Genome evolution: Horizontal movements in the fungi}.
\newblock \emph{Current Biology} \textbf{21}, R166--R168.
\newblock ISSN 09609822.
\newblock (\doi{10.1016/j.cub.2011.01.028}).

\bibitem{Gibbons2013}
Gibbons, J.~G. \& Rokas, A., 2013 {The function and evolution of the
  Aspergillus genome}.
\newblock \emph{Trends in Microbiology} \textbf{21}, 14--22.
\newblock ISSN 0966842X.
\newblock (\doi{10.1016/j.tim.2012.09.005}).

\bibitem{Hirt2015}
Hirt, R.~P., Alsmark, C. \& Embley, T.~M., 2015 {Lateral gene transfers and the
  origins of the eukaryote proteome: a view from microbial parasites}.
\newblock \emph{Current Opinion in Microbiology} \textbf{23}, 155--162.
\newblock ISSN 13695274.
\newblock (\doi{10.1016/j.mib.2014.11.018}).

\bibitem{Tomitani2006}
Tomitani, A., Knoll, A.~H., Cavanaugh, C.~M. \& Ohno, T., 2006 {The
  evolutionary diversification of cyanobacteria: molecular-phylogenetic and
  paleontological perspectives}.
\newblock \emph{Proceedings of the National Academy of Sciences of the United
  States of America} \textbf{103}, 5442--5447.
\newblock ISSN 0027-8424.

\bibitem{Kaufman2007}
Kaufman, A.~J., Johnston, D.~T., Farquhar, J., Masterson, A.~L., Lyons, T.~W.,
  Bates, S., Anbar, A.~D., Arnold, G.~L., Garvin, J. \& Buick, R., 2007 {Late
  Archean Biospheric Oxygenation and Atmospheric Evolution}.
\newblock \emph{Science} \textbf{317}, 1900--1903.
\newblock ISSN 1095-9203.

\bibitem{Edgar2004}
Edgar, R.~C., 2004 {MUSCLE: a multiple sequence alignment method with reduced
  time and space complexity.}
\newblock \emph{BMC Bioinformatics} \textbf{5}, 113.

\bibitem{Castresana2000}
Castresana, J., 2000 {Selection of conserved blocks from multiple alignments
  for their use in phylogenetic analysis}.
\newblock \emph{Molecular Biology and Evolution} \textbf{17}, 540--552.
\newblock ISSN 1537-1719.

\bibitem{SQ2008}
Le, S.~Q. \& Gascuel, O., 2008 {An improved general amino acid replacement
  matrix}.
\newblock \emph{Molecular Biology and Evolution} \textbf{25}, 1307--20.
\newblock ISSN 1537-1719.
\newblock (\doi{10.1093/molbev/msn067}).

\bibitem{Yang1994}
Yang, Z., 1994 {Maximum likelihood phylogenetic estimation from DNA sequences
  with variable rates over sites: approximate methods}.
\newblock \emph{Journal of Molecular Evolution} \textbf{39}, 306--314.
\newblock ISSN 1432-1432.

\bibitem{N2004}
Lartillot, N. \& Philippe, H., 2004 {A Bayesian mixture model for across-site
  heterogeneities in the amino-acid replacement process}.
\newblock \emph{Molecular Biology and Evolution} \textbf{21}, 1095--1109.
\newblock ISSN 07374038.
\newblock (\doi{10.1093/molbev/msh112}).

\bibitem{Fitzpatrick2006}
Fitzpatrick, D.~A., Logue, M.~E., Stajich, J.~E. \& Butler, G., 2006 {A fungal
  phylogeny based on 42 complete genomes derived from supertree and combined
  gene analysis.}
\newblock \emph{BMC evolutionary biology} \textbf{6}, 99.
\newblock ISSN 1471-2148.
\newblock (\doi{10.1186/1471-2148-6-99}).

\bibitem{Bansal2013}
Bansal, M.~S., Banay, G., Harlow, T.~J., Gogarten, J.~P. \& Shamir, R., 2013
  {Systematic inference of highways of horizontal gene transfer in
  prokaryotes}.
\newblock \emph{Bioinformatics} \textbf{29}, 571--579.
\newblock ISSN 13674803.
\newblock (\doi{10.1093/bioinformatics/btt021}).

\end{thebibliography}

\end{document}